\documentclass{article}
\usepackage{spconf,amsmath,graphicx}
\usepackage{cite}

\title{THE FIRST COMPREHENSIVE DATASET WITH MULTIPLE DISTORTION TYPES FOR VISUAL JUST-NOTICEABLE DIFFERENCES}
\name{Yaxuan Liu$^{a, \star}$ \qquad Jian Jin$^{b, \star}$ \qquad Yuan Xue$^{c}$ \qquad Weisi Lin$^{b}$ 
\thanks{© 20XX IEEE. Personal use of this material is permitted. Permission from IEEE must be obtained for all other uses, in any current or future media, including reprinting/republishing this material for advertising or promotional purposes, creating new collective works, for resale or redistribution to servers or lists, or reuse of any copyrighted component of this work in other works. $^{\star}$Equal contributions
}
}
\address{$^{a}$ Harbin Engineering University, College of Intelligent Systems Science and Engineering, Harbin, China \\
      $^{b}$ Nanyang Technological University, School of Computer Science and Engineering, Singapore \\
      $^{c}$ Fudan University, School of Software, Shanghai, China}
\begin{document}
%
\maketitle
\begin{abstract}
Recently, with the development of deep learning, a number of Just Noticeable Difference (JND) datasets have been built for JND modeling. However, all the existing JND datasets only label the JND points based on the level of compression distortion. JND models learned from such datasets can only be used for image/video compression. Hence, a generalized JND modeling should take more kinds of distortion types into account. To benefit JND modeling, this work establishes a generalized JND dataset with a coarse-to-fine JND selection, which contains 106 source images and 1,642 JND maps, covering 25 distortion types. To this end, we proposed a coarse JND candidate selection scheme to select the distorted images from the existing Image Quality Assessment (IQA) datasets as JND candidates instead of generating JND maps ourselves. Then, a fine JND selection is carried out on the JND candidates with a crowdsourced subjective assessment.
\end{abstract}
\section{Introduction}
\label{sec:intro}

JND is a metric for assessing the visual redundancy of the HVS, and it has been widely applied to computer vision and multimedia signal processing applications such as perceptual image and video compression \cite{wu2013perceptual} \cite{kim2015hevc}, image enhancement \cite{cheng2018performance}, watermarking \cite{li2019orientation}, and so on. JND modeling, which has been studied for many years, tries to precisely predict the visual redundancy of the HVS for a given visual content. Traditional JND models \cite{chou1995perceptually} \cite{wu2017enhanced} tried to predict the JND threshold for each pixel or each coefficient of the sub-bands based on the features of the HVS and their associated maskings. 
\begin{figure}
     \centering
     \includegraphics[width=0.45\textwidth]{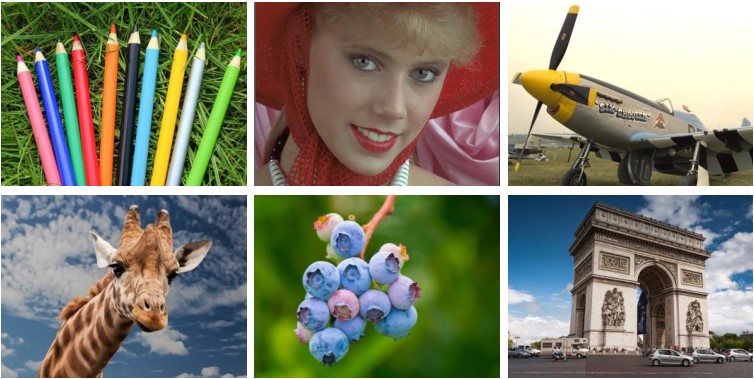}
     \caption{Illustration of some source images in our dataset}
     \label{fig5}
     \vspace{-0.5cm}
\end{figure}
Recently, a few works tried to extend the JND concept to a picture level and proposed the picture-wise JND. To utilize the powerful deep learning techniques in the JND modeling, lots of JND datasets \cite{jin2016statistical,shen2020jnd,lin2022large} were built. Jin et al. \cite{jin2016statistical} created the MCL-JCI dataset, which consists of 5,000 corresponding JPEG distorted versions with quality factors (QF) ranging from 1 to 100 and 50 source images in the $1920 \times 1080$ resolution. They conducted subjective quality assessment tests involving 150 participants watching the source image and a compressed image side by side on a TV. JND samples were gathered from 30 individuals for a specific image. Then, Shen et al. \cite{shen2020jnd} used 202 source images and 7,878 encoded versions of those images with a resolution of $1920 \times 1080$ to create a JND dataset based on the impending video coding standard Versatile Video Coding (VVC). Each source image was compressed by VTM5.0 coding with quantization parameters (QP) ranging from 13 to 51. The subjective tests were conducted in a carefully monitored lab setting. 20 subjects evaluated 20 PJND samples for each source image. Lin et al. \cite{lin2022large} expanded the datasets mentioned above and built a large PJND dataset called KonJND-1k. They conducted subjective JND assessment studies using the flicker test via crowdsourcing instead of in the laboratory environment. This dataset contains 1,008 source images as well as distorted versions created using JPEG and BPG compression. The study involved 503 workers in total, resulting in 61,030 PJND samples and an average of 42 samples per source image. However, all of these JND datasets in existence only took into consideration the impacts of compression distortion without considering the effects of the other distortion types. That is, the JND points were labeled based on the levels of compression distortion, \emph{e.g.}, Quality Factor (QF), Quantization Parameter (QP), \emph{etc}. Therefore, JND models learned from such kinds of datasets were limited to compression-relevant applications. In order to predict a JND model so that it has more wide applications, establishing a generalized JND dataset that covers all existing distortion types is necessary and significant. However, the challenging part is how to obtain distorted images that cover all distortion types with different distortion levels so that JND maps can be selected from them. 

Recently, many IQA datasets \cite{lin2019kadid,larson2010most,ponomarenko2009tid2008,ponomarenko2015image} have been established, which contain source images and their associated distortion images as well as subjective scores, \emph{e.g.}, Mean Opinion Scores (MOS). Commonly, their distortion images cover various distortion types under various distortion levels. In view of this, we can directly pick out the JND maps from the distorted images in the IQA datasets. On the one hand, the generation of distortion images that cover all distortion types with different distortion levels can be saved. On the other hand, as the MOS reflects the perceptual quality of distorted images, it can be used for selecting JND maps. Therefore, we directly select the JND maps from IQA datasets with a proposed coarse-to-fine JND map selection scheme, which contains two main steps: 1) coarse JND candidate selection with MOS as the threshold and 2) fine JND map selection with crowdsourced subjective assessment. The first step helps us make a fast JND candidate selection from huge distorted images, which can save time and human resources in subsequent subjective assessment. The second step subjectively confirms the final JND maps from the selected JND candidates in the first step by the subjects. Finally, we establish a generalized JND dataset, containing 106 source images (part of them are shown in Fig. \ref{fig5}) and 1,642 JND maps, with 25 distortion types being involved. It should be mentioned that this is the first JND dataset that covers various distortion types.

\section{Coarse-to-Fine JND Selection}
\label{sec:jndselec}

\subsection{Coarse JND Candidate Selection}
\label{ssec:coarseselec}

To select JND images from these datasets, we use the MOS of distorted images as the preliminary selection criterion, since the MOS is a subjective viewing score, reflecting the perceptual quality of distorted images among all the subjects. Commonly, the higher the distortion image quality, the smaller the difference between it and its associated source image, and vice versa. Therefore, we can use MOS to select distorted images with high perceptual quality as the JND maps, so as to reduce the amount of data in subsequent subjective tests. The involved IQA datasets in this work include TID2013 \cite{ponomarenko2015image}, TID2008 \cite{ponomarenko2009tid2008}, and KADID-10k \cite{lin2019kadid}, which have a total of 14,825 distortion images from 106 source images, and 39 distortion types, each with 4 or 5 distortion levels. 

\begin{figure}[htbp]
\centerline{\includegraphics[width=0.45\textwidth]{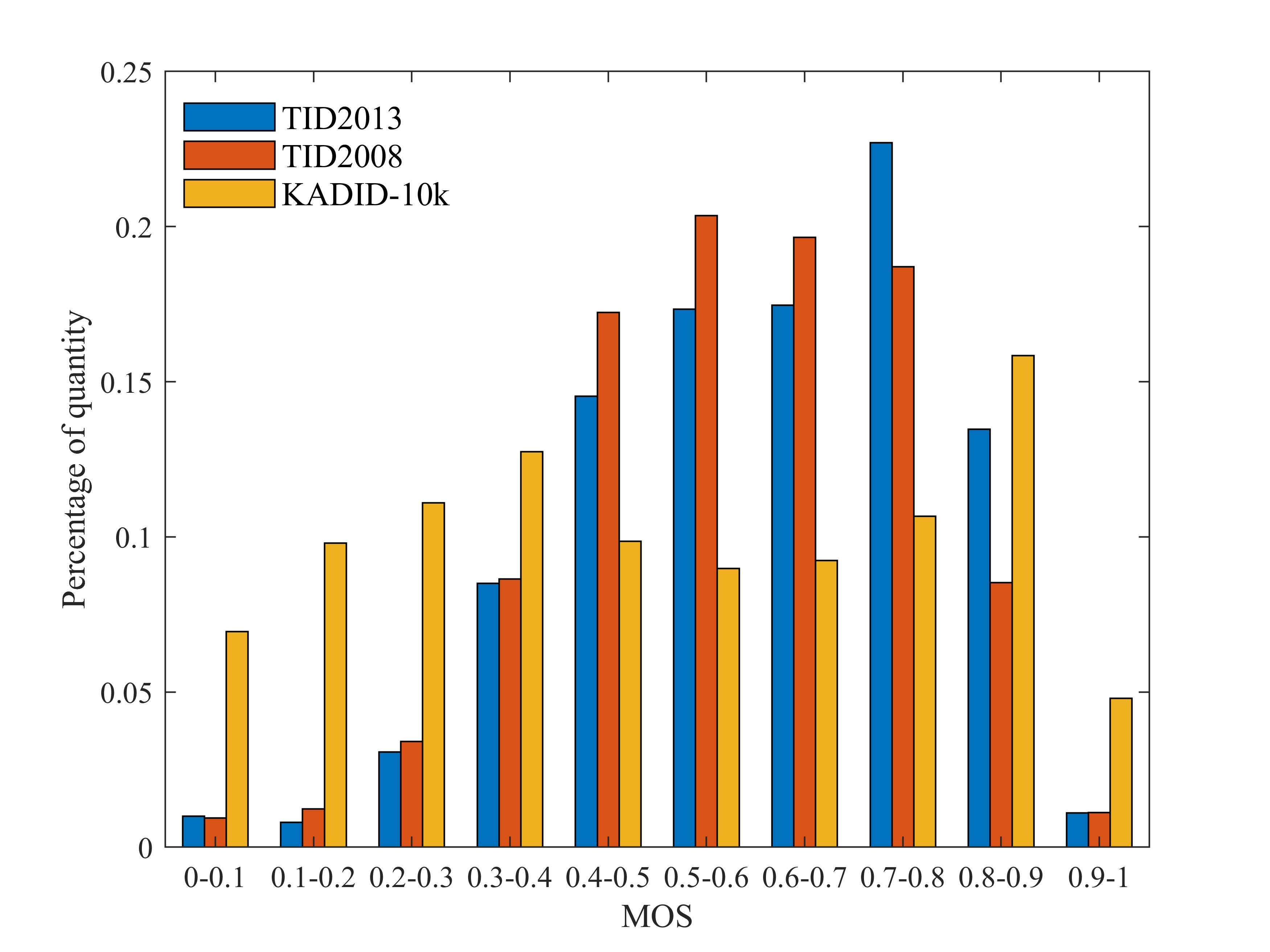}}
\caption{Histogram of the distribution of normalized MOS in TID2008, TID2013, and KADID-10k datasets, which shows that different IQA datasets have different MOS distributions.}
\label{fig1}
\end{figure}

Since different IQA datasets use different algorithms to subjectively evaluate distortion images, the range of the obtained MOS values is different. In view of this, we first normalize the MOS values. This process can be formulated as 
\begin{equation}
Q=\frac{P-Min}{Max-Min},
\end{equation}
where $P$ represents the MOS value of the distortion image in a certain IQA dataset, $Min$ and $Max$ are the minimum and maximum values of the MOS in the current IQA dataset. $Q$ represents the normalized MOS. The range of the normalized MOS is [0,1]. Fig. \ref{fig1} shows the MOS distribution of each IQA dataset after normalization, and it can be seen that the MOS distribution in different datasets is different. Among them, the MOS distribution of TID2013 and TID2008 conforms to the normal distribution, while the KADID-10k dataset is relatively evenly distributed in each interval. To determine a threshold that can distinguish between JND and non-JND maps, we randomly sample 10\%-20\% of the source images and all their corresponding distorted images from each IQA dataset and then pick out JND maps from them through a subjective viewing test by well-trained and experienced subjects. We only sample 10\%-20\% of the whole images because the number of images is huge and we don't need to ensure the high accuracy of the threshold due to the fine JND map selection after. As a result, 112 JND maps are confirmed out of 2,190 distorted images with MOS above 0.7. To determine the MOS threshold, we calculate the average MOS of the confirmed JND images for each image dataset and then reduce the calculated average MOS by about 5\% so that more JND-eligible maps can be involved. Then, we obtain the MOS thresholds 0.855 for the KADID-10k dataset, and 0.795 for the TID2013 and TID2008 datasets, respectively. Afterward, we use the MOS thresholds above to pick out images from their corresponding datasets as the JND candidates. In this way, we coarsely obtain 1,962 JND candidates.

\subsection{Fine JND Map Selection}
\label{ssec:fineselec}

After coarse JND selection, there are still some JND candidates that can be noticed the differences compared with their associated source images. In view of this, we conduct a crowdsourced subjective assessment of the JND candidates to achieve fine JND selection by abandoning non-JND maps. The crowdsourced subjective assessment was conducted on Amazon Mechanical Turk (AMT). On the AMT platform, requesters—companies, organizations, or individuals—create and submit tasks requiring human intelligence for workers, who may be paid for each task successfully completed.

In this work, we adopt the flicker test as used in \cite{iso2015}, where the JND candidate and its associated source image are toggled back and forth successively at a frequency of 8 Hz. Here, we set three options for subjects: obvious flicker, slight flicker, and no flicker. The definitions of the three options are as follows. The obvious flicker indicates that the flicker is intense, affecting the entire image. The slight flicker means that slight changes can be seen in the image, such as a subtle variation in the image brightness. These changes can resemble a mild film grain in a static scene of a movie and are often localized to small regions of the image. No flicker represents the image is static, there is no visible change in the image, no matter how small it is.

According to the definition of JND as aforementioned in Sec. \ref{sec:intro}, no flicker and slight flicker JND candidates are selected as the final JND maps. Here, we randomly divide all the 1962 JND candidates into 103 groups. Each group contains 19 or 20 JND candidates, which are assessed by 30 workers (subjects). A total of $1962\times30$ results are obtained.

Considering that the screens used by the workers may have different sizes and resolutions, resulting in different physical sizes of pictures displayed on the screen, we carry out the calibration process in \cite{lin2022large} to make the image appear the same physical size on all workers' computers. That is, workers must prepare a card the size of a credit card ($85mm \times 53.98 mm$) and change the size of a frame on the screen to fit the card. By computing the Logical Pixel Density (LPD) of the display in Pixels Per Inch (PPI), we can estimate the physical size of the display. For more details refer to \cite{lin2022large}. Following the calibration, we instruct the workers to change their viewing distance to $30cm$ in accordance with the trigonometric calculation \cite{rohrschneider2004determination}, \cite{ehinger2017humans} and ISO standard \cite{iso2010ergonomics}.

After passing a training phase, the workers will enter the assessment phase. In the assessment phase, 19 or 20 images are set for each task, and the workers choose the most suitable option from the three options of no flicker, slight flicker, and obvious flicker, as aforementioned.

\section{Results Processing and Analysis}
\label{sec:respross}

\subsection{Outlier Removal}
\label{ssec:outremove}

To get effective results, we need to remove outliers in the collected results. Since we can not guarantee that all workers complete the test with seriousness, for example, some workers may randomly choose an option during the test, or workers may submit unreliable answers because they have been tested for a long time. In this paper, we adopt the result processing method mentioned in the ITU-R Recommendation BT.500 \cite{series2012methodology} to remove unreliable results. We use the scores 1, 2, and 3 to represent no flicker, slight flicker, and obvious flicker respectively. Then, we calculate the mean and standard deviation of the scores rated by all the subjects for each image as follows 
\begin{equation}
\bar{u}_j=\frac{1}{N}\sum_{i=1}^N{u_{ij}},
\end{equation}
\begin{equation}
S_j=\sqrt{\sum_{i=1}^N{\frac{\left( \bar{u}_j-u_{ij} \right) ^2}{\left( N-1 \right)}}}.
\end{equation}
$u_{ij}$ denotes the score rated by the $i$-th subject on the $j$-th image, where $i=1, 2, ..., N$ and $j=1, 2, ..., M$. $\bar{u}_j$ and $S_j$ are the mean and standard deviation of scores rated by all the subjects for the $j$-th image.
Subsequently, we adopt the confidence interval in \cite{series2012methodology} for outlier removal, that is  

\begin{equation}
   \left[ \bar{u}_j-\delta _j,\bar{u}_j+\delta _j \right], \text{where}\ \delta _j=1.96 \frac{S_j}{\sqrt{N}}. 
\end{equation}

$[ \cdot ]$ is rounding operation. $u_{ij}$ outside such an interval is considered as an outlier. After removing the outliers, we re-calculate the average score of each image. We regard the image with an average score of less than 2.5 as a JND map.

\subsection{Result Analysis}
\label{ssec:resanaly}

\begin{table}[t]
\caption{Numbers of JND Maps and The Ratio of The Number of JND Maps to The Total Samples for Different Distortion Types}
\begin{center}
\begin{tabular}{|l|c|c|}
\hline \textbf{No.  Distortion types} & \textbf{Num.} & \textbf{Ratio} \\
\hline  01  Additive Gaussian noise & 19 & 0.030 \\
        02  Additive noise in color components & 115 & 0.183 \\
        03  Masked noise & 50 & 0.222 \\
        04  High frequency noise & 38 & 0.167\\
        05  Denoise & 50 & 0.079 \\
        06  Multiplicative noise & 55 & 0.104 \\
        07  Comfort noise & 18 & 0.144 \\
        08  Motion blur & 91 & 0.225 \\
        09  Gaussian blur & 128 & 0.203 \\
        10  Color diffusion & 60 & 0.148 \\
        11  Color shift & 25 & 0.062\\
        12  Color quantization & 25 & 0.047 \\
        13  Color saturation & 70 & 0.132 \\
        14  JPEG & 95 & 0.151\\
        15  JPEG transmission errors & 26 & 0.116 \\
        16  JPEG2000 & 123 & 0.195 \\
        17  Darken & 138 & 0.341 \\
        18  Brighten & 68 & 0.168 \\
        19  Mean shift & 233 & 0.370 \\
        20  Jitter & 40 & 0.099 \\
        21  Non-eccentricity patch & 25 & 0.040 \\
        22  Quantization & 36 & 0.089 \\
        23  Contrast change & 54 & 0.086 \\
        24  Sparse sampling and reconstruction & 13 & 0.104 \\
        25  Chromatic aberrations & 47 & 0.376 \\
\hline
\end{tabular}
\label{tab1}
\end{center}
\end{table}
As a result, we select a total of 1,642 distortion images (corresponding to 106 source images) as the JND maps, which contain 25 distortion types. Table \ref{tab1} shows the 25 distortion types, their number of JND maps, and the ratio of the number of JND maps to the total number of samples. From the table, we can see that the number of JND maps with different distortion types varies greatly. There are two reasons to explain this. First, since we select JND maps from three IQA datasets KADID-10k, TID2008 and TID2013, some distortion types are not contained by all three datasets such as Comfort noise and Sparse sampling and reconstruction, which causes different numbers of samples for different distortion types. The distortion type with a small number of samples probably has fewer JND maps and vice versa. Second, the HVS has different sensibilities to different types of distortion \cite{wu2017enhanced}. For the distortion type with a high sensibility of the HVS, even low-level distortion may be perceived by the HVS and this will make fewer JND maps can be obtained from the IQA datasets, and vice versa. Therefore, the distortion with a higher sensibility to the HVS, the fewer JND maps we obtained from the samples with such type of distortion. Besides, for each distortion type, we also calculate the ratio of the number of JND maps and total samples. The difference in such ratios reflects a certain extent to which types of distortion are more easily sensitive to the HVS.

\begin{figure}[t]
\centerline{\includegraphics[width=0.47\textwidth]{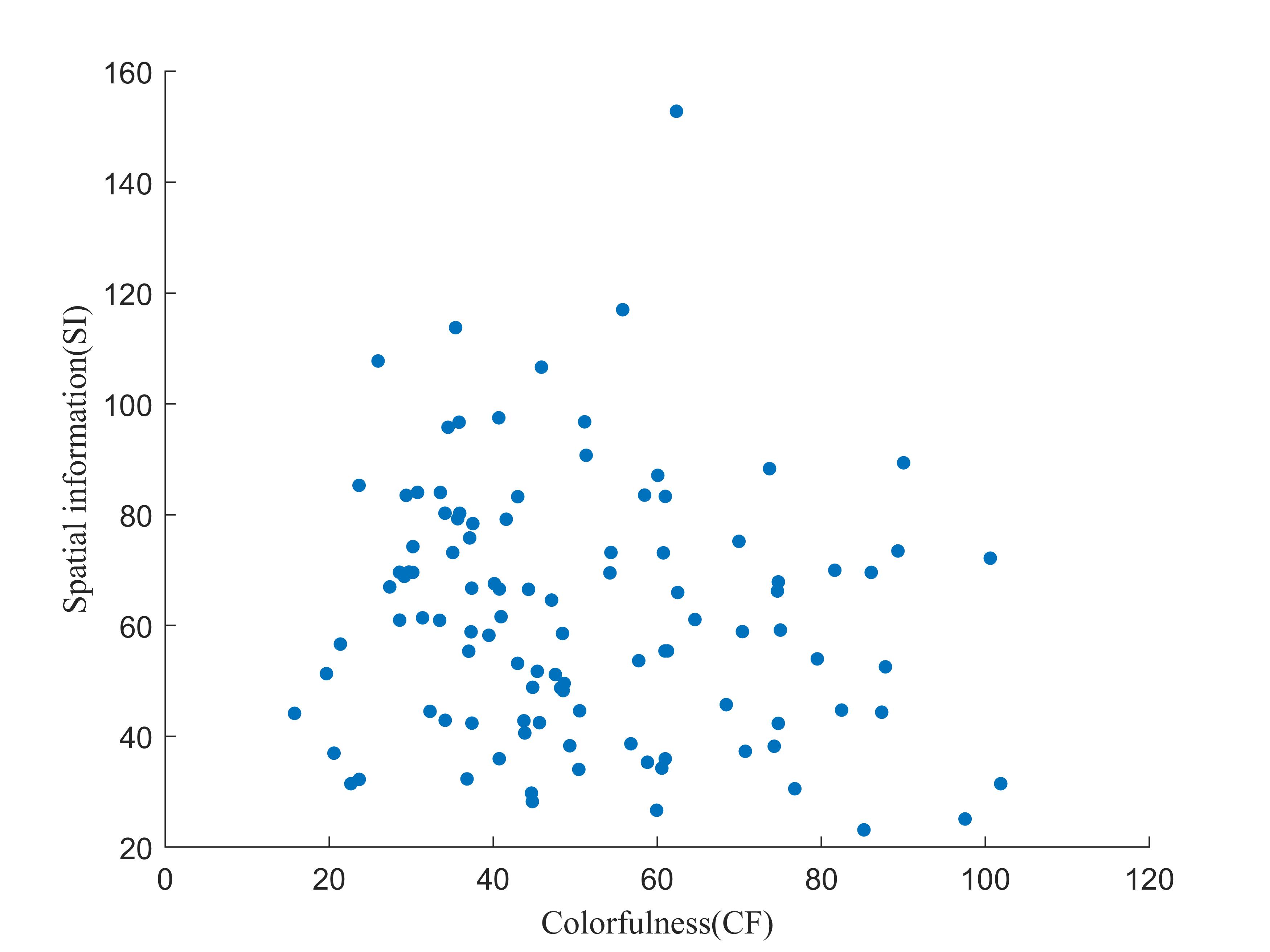}}
\caption{Distribution of spatial information and colorfulness of source images in our dataset.}
\label{fig3}
\end{figure}

A good JND dataset tends to contain a wide variety of visual content. To demonstrate the diversity of the JND maps in our dataset, we calculated the spatial information (SI) \cite{itu1999subjective} and the colorfulness (CF) \cite{winkler2012analysis} of 106 source images, as shown in Fig. \ref{fig3}. The spatial information reflects the spatial complexity of the image, and the colorfulness reflects the richness of the color of the image. In Fig. \ref{fig3}, we can see that the selected source images do have wide coverage in this plot. Furthermore, we divide the 106 source images into 8 semantic categories, as shown in Table \ref{tab2}. 

\begin{table}[htbp]
\caption{Number of Source Images in Each Semantic Category}
\begin{center}
\begin{tabular}{|c|c|}

\hline \textbf{Content} & \textbf{Number} \\
\hline       People & 12 \\
\hline       Animals & 16 \\
\hline       Plants & 12 \\
\hline       Objects & 21 \\
\hline       Landscape & 9 \\
\hline       Outdoor & 15 \\
\hline       Building & 12 \\
\hline       Nature & 9 \\
\hline
\end{tabular}
\label{tab2}
\end{center}
\end{table}

\section{Conclusions}
\label{sec:conclusion}

A JND dataset containing 106 source images and 1,642 JND maps with 25 distortion types involved has been established in this work. To have such a dataset, we coarsely select the JND candidates from IQA datasets through the MOS thresholds, and then subjectively select the final JND map from the JND candidates by conducting crowdsourced subjective assessments, which can greatly save time and human resources spent generating distorted images by ourselves.
This is the first JND dataset covering multiple distortion types, which can be used to learn JND models for more distortion types, not just limited to compression distortion types contained in existing JND datasets, and thus have a wider range of applications. For example, we can use it to build a JND model to optimize the image transmission process and reduce the image distortion caused by packet loss during data transmission. Furthermore, considering that pictures on social media now contain multiple distortions, this dataset can also be used to build a more generalized JND model, which can well simulate the perception of human vision for any type of distortion, thus revealing the characteristics of the human visual system and improving the quality of images on social media sites. Besides, our dataset has strong expansibility. In our future work, we will extend this dataset by applying the proposed coarse-to-fine JND selection scheme to more IQA datasets such as PIPAL \cite{jinjin2020pipal}, CSIQ \cite{larson2010most}, and KADIS-700k \cite{lin2020deepfl}.


\bibliographystyle{IEEEtran}
\bibliography{refs}

\end{document}